\documentclass[english,showpacs,aps,superscriptaddress]{revtex4}
\usepackage{pslatex}
\usepackage[T1]{fontenc}
\usepackage{color}
\usepackage{graphicx}

\makeatletter

\usepackage{babel}
\makeatother
\begin{document}

\title{Local atomic and electronic structure in LaMnO$_{3}$ across the
orbital ordering transition }

\author{Raquel A. Souza}

\affiliation{Laboratório Nacional de Luz Síncrotron - LNLS, P.O. Box 6192, 13084-971,
Campinas, São Paulo, Brazil}

\affiliation{Instituto de Física Gleb Wataghin, IFGW - UNICAMP, Campinas, SP,
Brazil}

\author{Narcizo M. Souza-Neto}

\affiliation{Laboratório Nacional de Luz Síncrotron - LNLS, P.O. Box 6192, 13084-971,
Campinas, São Paulo, Brazil}

\affiliation{Dept. de Física dos Materiais e Mecânica, DFMT-IF-USP, São Paulo,
SP, Brazil}

\author{Aline Y. Ramos}

\email{aramos@lnls.br}

\affiliation{Laboratório Nacional de Luz Síncrotron - LNLS, P.O. Box 6192, 13084-971,
Campinas, São Paulo, Brazil}

\affiliation{Laboratoire de Minéralogie-Cristallographie de Paris, LMCP -UMR 7590
-CNRS, Paris, France}

\author{Hélio C. N. Tolentino}

\affiliation{Laboratório Nacional de Luz Síncrotron - LNLS, P.O. Box 6192, 13084-971,
Campinas, São Paulo, Brazil}

\author{Eduardo Granado}

\affiliation{Laboratório Nacional de Luz Síncrotron - LNLS, P.O. Box 6192, 13084-971,
Campinas, São Paulo, Brazil}

\affiliation{Instituto de Física Gleb Wataghin, IFGW - UNICAMP, Campinas, SP,
Brazil}

\date{\today{}}

\begin{abstract}
The local atomic disorder and the electronic structure in the environment
of manganese atoms in LaMnO$_{3}$ has been studied by x-ray absorption
spectroscopy, over a temperature range (300K to 870K) covering the
orbital ordering transition ($\sim$710K). The Mn-O distances splitting
into short and long bonds (1.95 and 2.15\AA) is kept across the transition
temperature, so that the MnO$_{6}$ octahedra remain locally Jahn-Teller
distorted. Discontinuities in the Mn local structure are identified
in the extended x-ray fine structure spectra at this temperature,
associated to a reduction of the disorder in the super-exchange angle
and to the removal of the anisotropy in the radial disorder within
the coordination shell. Subtle changes in the electronic local structure
also take place at the Mn site at the transition temperature. The
near edge spectra show a small drop of the Mn 4$p-$hole count and
a small enhancement in the pre-edge structures at the transition temperature.
These features are associated to an increase of the covalence of the
Mn-O bonds. Our results shed light on the local electronic and structural
phenomena in a model of order-disorder transition, where the cooperative
distortion is overcome by the thermal disorder.
\end{abstract}

\pacs{75.47.lx, 61.10.Ht, 71.90.+q}

\maketitle

\section{Introduction}

The lanthanum manganite and its doped perovskite alloys La$_{1-x}$$A$$_{x}$MnO$_{3}$
(with $A$= Ca, Sr, Ba) have attracted much attention in the last
decade, in large part due to the potential applications of the so
called colossal magnetoresistance\cite{Millis-N98}. The unusual physical
properties of these compounds arise from intricate interrelations
between spin, charge and local structure. In spite of the large experimental
and theoretical efforts of the scientific community many of these
interrelations remain not elucidated. In the LaMnO$_{3}$ compound
itself many questions remain to be addressed. At room temperature
LaMnO$_{3}$ is an antiferromagnetic semiconductor and crystallize
in an orthorhombic variant of the cubic perovskite structure space
group $Pbnm$. The MnO$_{6}$ octahedra in LaMnO$_{3}$ are distorted
due to the Jahn-Teller ($JT)$ effects of the Mn$^{3+}$($t_{2g}^{3}e_{g}^{1}$)
and the $Mn-O$ distances are split into two groups. The manganese
atoms are coupled ferromagnetically in the ab plane and antiferromagnetically
along the c axis and an orbital ordering takes place. In the basal
ab plane long and short $Mn-O$ bonds alternates. The apical and basal
short bonds have different length ($\approx1.91$\AA ~and $\approx1.97$\AA~respectively),
however this additional splitting is small for local probes such as
real space high resolution diffraction \cite{Proffen-PRB99,Billinge-PRB00,Louca-PRB00}
and extended x-ray absorption fine structure (EXAFS)\cite{Booth-PRB98,Subias-PRB98,Shibata-PRB03}.
The local radial distribution is thus seen as made of a single distance
at $(Mn-O)_{s}\approx1.94$\AA, separated from the long distance
long bond $(Mn-O)_{l}\approx2.15$\AA. LaMnO$_{3}$ undergoes a transition
at $T^{*}$ $\approx$710-750 K from the $JT$ distorted orthorhombic
phase to a high temperature nearly cubic phase \cite{Rodriguez-Carvajal-PRB98}.
The space group $Pbnm$ remains the same in both phases, but the cell
distortion is nearly removed and the orbital ordering disappears in
the high temperature phase. The transition is accompanied by abrupt
changes in the electrical resistivity, thermoelectric power and Weiss
constant \cite{Zhou-PRB99}.

At the local scale two scenarios may be proposed: symmetrization of
the distorted octrahedra MnO$_{6}$~or an order-disorder transition,
where the local distortion is maintained. From thermodynamic calculations
it has been shown \cite{Millis-PRB96} that the thermal energy at
$T^{*}$ is small with respect to the gain associated to the lift
of the $e_{g}$ levels degeneracy, so that this transition will hardly
correspond to a removal of the octahedral distortion. The transition
would happen as an order-disorder transition in the sequence of the
distorted octahedra. Experimentally, two early works from Raman spectroscopy
\cite{Granado-PRB00} and from the analysis of the overall disorder
in the EXAFS data across $T^{*}$ \cite{Araya-Rodriguez-JMMM01},
supported the hypothesis of the upkeep of the local $JT$ distortion.
On the other hand the hypothesis that the local $JT$ distortion may
vanish and the $Mn-O$ distances collapse, was sustained by the work
in thin LaMnO$_{3}$~films \cite{Song-PRB02}. The two distances
found by the fitting procedure \cite{Song-PRB02} collapse into a
single one above a given temperature, in a similar way as the cell
parameters. More recently, Sanchez et al.\cite{Sanchez-PRL03} have
reached opposite conclusions from their x- ray absorption study in
LaMnO$_{3}$ polycrystals. 

There have been various works using X-ray absorption spectroscopy
(XAS), shedding light on the nature of the Jahn-Teller distortion,
electronic states, thermal behavior and disorder in manganites\cite{Booth-PRB98,Subias-PRB98,Shibata-PRB03,Araya-Rodriguez-JMMM01,Sanchez-PRL03,Tyson-PRB96,Lanzara-PRL98,Croft-PRB97,Bridges-PRB00,Bridges-PRB01a,Qian-PRB00,Qian-PRB03}.
The thermal and the structural disorder enter equivalently in the
theory and in the data analysis where the overall disorder is usually
expressed by a Debye Waller-like factor (DWF). The thermal disorder
depends on the dynamical properties of the lattice. It presents a
smooth continuous increase with the temperature that can be modeled
using models of correlated vibrations \cite{Sevillano-PRB79}. The
structural local disorder accounts for bonds length dispersion and
site distortions that, in the absence of structural transition, are
temperature independent. At high temperatures, the EXAFS signal is
strongly damped by the thermal disorder. The limited available analysis
range reduces the R-resolution and closely correlated fitting parameters
cannot always been unambiguously resolved\cite{Booth-PRB98}. In addition,
the structural disorder associated to the long and the short $Mn-O$
bonds are significantly different \cite{Shibata-PRB03,Subias-PRB98}
and the thermal behavior of the DWF may not be correlated. To avoid
misfits in the choice of fixed or correlated parameters, qualitative
methods using an alternative way of data handling may be useful, prior
to the application of the fitting procedure. The phase derivative
method, based on the analysis of the beat stemming from the combination
of two close frequency sinusoids, has proved in several complex systems
its usefulness as a diagnostic tool\cite{Martens-PRL77,Jaffres-PRB00,Jiang-PRB91}.
This method, by characterizing the occurrence of a distance splitting,
should be especially useful to identify possible collapse of this
splitting as the temperature is increased. With the support of this
evidence, the EXAFS fitting analysis should then be specially focused
on the behavior of the disorder related Debye Waller-like term.

X-ray absorption near edge spectroscopy (XANES) has been shown to
be very sensitive to the various degrees of freedom governing the
electronic properties of manganite compounds, such as local atomic
distortion, charge transfer in the $Mn-O$ bond, or local magnetic
ordering \cite{Booth-PRB98,Subias-PRB98,Shibata-PRB03,Croft-PRB97,Bridges-PRB00,Bridges-PRB01a,Qian-PRB00,Qian-PRB03}.
The temperature dependence of XANES spectra in LaMnO$_{3}$ has been
the object of some reports in the literature\cite{Bridges-PRB00,Bridges-PRB01a,Qian-PRB00},
but most of them are related to the thermal behavior of this compound
around the magnetic ordering temperature ($\sim$140K). It can be
expected that temperature dependence of XANES spectra across the orthorhombic
to cubic transition may shed more light on the modification in the
local atomic and electronic structure at the site of the Mn$^{3+}$
ions.

We address here the issue of the temperature dependence of the local
distortion around Mn$^{3+}$ ions in LaMnO$_{3}$ with a complete
set of EXAFS and XANES data, collected below and above $T^{*}$. At
this temperature some discontinuities are observed in both EXAFS and
XANES range data. In the EXAFS range, the Jahn-Teller distance splitting
still exists within the coordination shell even above the transition,
but modification in the disorder can be identified. At low temperature
the static disorder associated to the long bond $(Mn-O)_{l}$ is about
three times larger than the disorder associated to the short bond
$(Mn-O)_{s}$. As the temperature increases the overall disorder associated
to the short bond is rapidly dominated by the thermal contribution
whereas the structural term is still dominant for the long bond. Around
$T^{*}$ these terms become of the same order and the thermal disorder
is likely to be dominant for all the apical and basal bonds. We conclude
then that the tetragonal distortion is preserved across $T^{*}$,
in agreement with an order-disorder character for the transition.
We point out that this transition takes place when the relative displacement
of the atoms along the long and short bonds becomes of the same order,
turning the average radial disorder isotropic and allowing the bond
length to vary independently. In the XANES range, the weak but significant
discontinuities observed across $T^{*}$ are associated to an increase
of the covalent character of the $Mn-O$ bonds.

\section{Experimental}

The Mn-K edge EXAFS spectra were collected on a powder of LaMnO$_{3}$
synthesized by the solid state reaction method. The sample shows an
orthorhombic to cubic transition at $T^{*}$ $\sim710K$. Details
of the synthesis method, the structural characterization and magnetic
of the sample have been published elsewhere \cite{Granado-PRB00}.
The temperature dependent x-ray absorption measurements were performed
at LNLS D04B-XAS1 beamline \cite{Tolentino-JSR01} in the transmission
mode. A fine grained powder sample of LaMnO$_{3}$ was pressed between
two thin beryllium windows and placed into an electrical furnace with
low vacuum ($<10^{-3}$Torr). EXAFS and XANES data were collected
at increasing temperature from 300K to 870K. Then the temperature
has been decreased and the XAS spectra have been collected again at
some temperature points and compared to those previously collected.
Using this procedure we can certify that the sample had not been oxidized
during the high temperature measurements. In addition we verify with
this procedure the perfect reproducibility of the subtle features
in the XANES spectra. The EXAFS data were collected up to k$_{max}$=12.5\AA$^{-1}$.
This data range limits the minimum difference between two close distances
that can be resolved in the EXAFS analysis to the value ($\delta$R)$_{\textrm{min}}$=
0.13\AA. The two distances at 1.91\AA~ and 1.97\AA~are seen as
one unique distance $(Mn-O)_{s}\approx1.94$\AA, easily resolved
from the long distance $(Mn-O)_{l}\approx2.15$\AA ~($\delta$R=
0.21\AA). The experimental resolution for the XANES experiments was
about 1.5eV. The simultaneous collection of XANES metal foil reference
spectra enables the eventual correction of small energy shift due
to thermal effects in the optics of the beam line. After background
subtraction the spectra were normalized in the range 150-250eV above
the edge. The features in the XANES spectra can then be compared in
intensity, and in position with a sensitivity as low as 0.1eV. 

The EXAFS oscillations were extracted following standard procedure\cite{Koningsberger88}.
The signal corresponding to the oxygen coordination shell is selected
by Fourier filtering. A first analysis has been made using the phase
derivative $(PD)$ method. This method is based on the exploration
of the modulation in the EXAFS signal occurring when the contributions
of two close shells separated by $\delta R$ are combined. For close
shells with the same backscattering atom, this modulation results
in a minimum in the total amplitude and an inflexion in the total
phase of the EXAFS signal, at $k=k{}_{B}$ given by $k_{B}\sim\pi/2\delta R$\cite{Martens-PRL77,Jaffres-PRB00,Jiang-PRB91}.
The presence of a beat is widely used to identify the occurrence of
close shells. Due to the approximations involved, $\delta R_{B}=\pi/2k_{B}$
does not give exactly the bond length separation, however any relative
modification in this separation will be detected by a shift in the
beat position. The second analysis consisted in the application of
the conventional fitting procedure constraining the parameters using
the information obtained from the $PD$ method. In the fitting procedure,
the number of free parameters is limited to $n\approx7$, by the useful
range and the interval corresponding to the selected signal in the
real space.

\section{Results and discussion}

\begin{figure}
\includegraphics{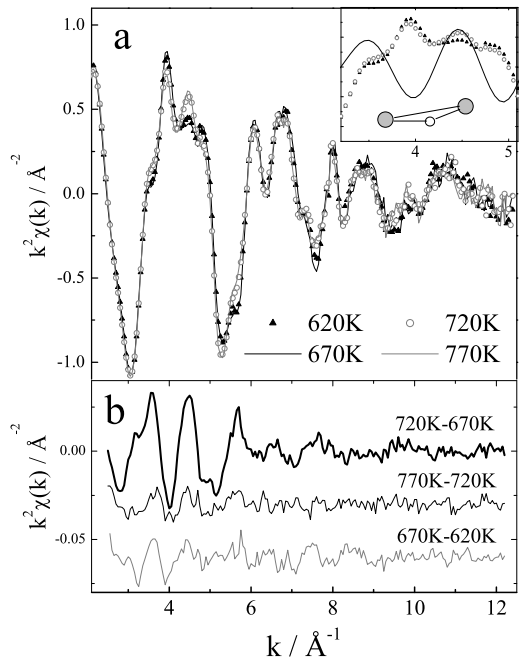}

\caption{\label{fig2-EXAFS} EXAFS signal (a) and difference spectra (b) below,
above and across the transition temperature $T^{*}$~.The inset is
a zoom of the first EXAFS oscillation showing the simulation of the
3-leg MS contribution, in phase with main additional structure above
the transition.}
\end{figure}

\begin{figure}
\includegraphics{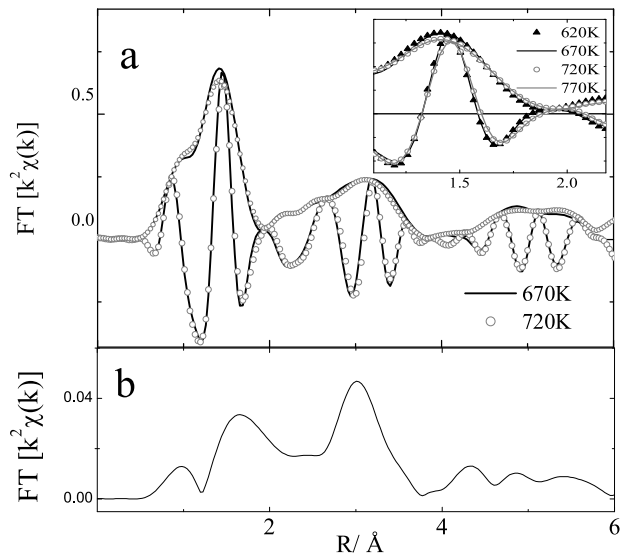}

\caption{\label{fig3-FT} (a) Fourier transform of the EXAFS signal across
$T^{*}$ (modulus and real part). Inset : detail of the first peak
corresponding to the coordination shell. Subtle but reproducible changes
are observed for temperature below and above the transition temperature.
(b) Modulus of the Fourier transform of the difference EXAFS signal
across $T^{*}$, showing the two differential contributions around
3 and 1.8 \AA.}
\end{figure}

The k$^{2}$-weighted $\chi$(k) spectra {[}$\chi(k)*k^{2}${]} for
selected temperatures below and above the cell ordering temperature
$T^{*}$($\sim$710K) are shown in the Fig.\ref{fig2-EXAFS}a. \textcolor{black}{On
the Fig.}\ref{fig2-EXAFS}b, \textcolor{black}{we report} the difference
EXAFS spectra for measurement at temperature differing by 50K, just
below, just above, and across the transition\textcolor{black}{.} Subtle
modifications are observed across $T^{*}$. The high frequency of
these modifications indicates that they are associated to higher shells
or multiple scattering (MS) effects. We would like to call attention
to the additional feature appearing around k =4.5\AA$^{-1}$in the
EXAFS spectra (Fig.\ref{fig2-EXAFS}a). This feature is common, with
different scale of intensities, to all reported EXAFS spectra of manganite
compounds, but is absent in LaMnO$_{3}$ below $T^{*}$ \cite{Booth-PRB98,Shibata-PRB03,Subias-PRB98,Sanchez-PRL03,Tyson-PRB96}.
Aiming to determine the exact origin of this contribution, we perform
\textit{ab initio} simulations in MS expansion, using the Feff7 code\cite{Rehr-JAChemSoc91,Rehr-PRL92}.
Different clusters were built, using the crystallographic structure
of \textit{Pbnm} and $R\overline{3}C$ doped compounds. A path by
path study allowed the identification of the additional contribution
above $T^{*}$ as being mainly due to 3-leg MS paths involving two
neighboring Mn atoms and their common oxygen (Fig.\ref{fig2-EXAFS}a,
inset). The contribution of this path increases with increasing the
superexchange angle (tilt angle) Mn-O-Mn, as well as with the reduction
of angular disorder. This contribution is clearly revealed, in the
Fourier transform (FT) of the EXAFS signal across $T^{*}$ (Fig. \ref{fig3-FT}).
Apart from other modifications, discussed below, a large peak around
3\AA~ corresponding to that MS contribution shows up in the difference
signal (Fig. \ref{fig3-FT}b). As neutron diffraction studies\cite{Rodriguez-Carvajal-PRB98}
do contain any report of an anomalous behavior in the average tilt
angle, we may conclude that the enhancement of the MS contribution
at $T^{*}$is more likely related to a reduction of the average angular
disorder. 

A closer view of the first peak in FT corresponding to the contribution
of the coordination shell (Fig.\ref{fig3-FT}a, inset) reveals small
but significant modifications occurring within the MnO$_{6}$ octahedra,
across the transition. The shape of the signal is not notabily affected,
but we observe a small drop in the height of the peak (at $\sim$1.3\AA)~and
a discontinuity in the thermal sequence, clearly seen in the real
FT component, in the high R side of the peak (at $\sim$1.8\AA)~,
associated to the long bond contribution \cite{Shibata-PRB03}. The
Fourier transform of the difference spectra across $T^{*}$(Fig.\ref{fig3-FT}b)
shows a peak around 1.8 \AA,~confirming the occurrence of a structural
modification within the coordination shell. As already pointed out,
the thermal damping may smooth and screen part of the real effects,
so that a thorough analysis is necessary to determinate the extent
of modification of the local atomic organization.

\begin{figure}
\includegraphics{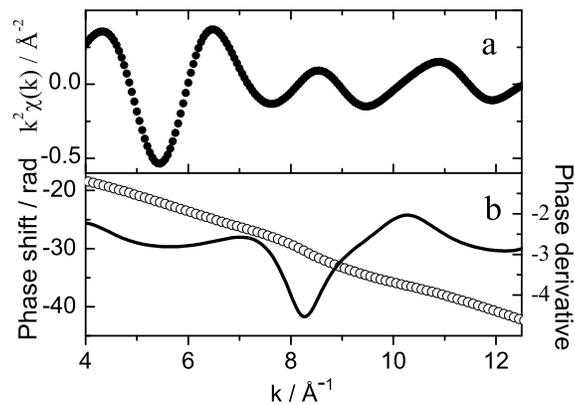}

\caption{\label{fig4-Phase} a. Back Fourier transformed signal of the coordination
shell at room temperature. b.Total phase shift and derivative: the
minimum of the curve defines the beat position $k_{B}$. }
\end{figure}

\begin{figure}
\includegraphics{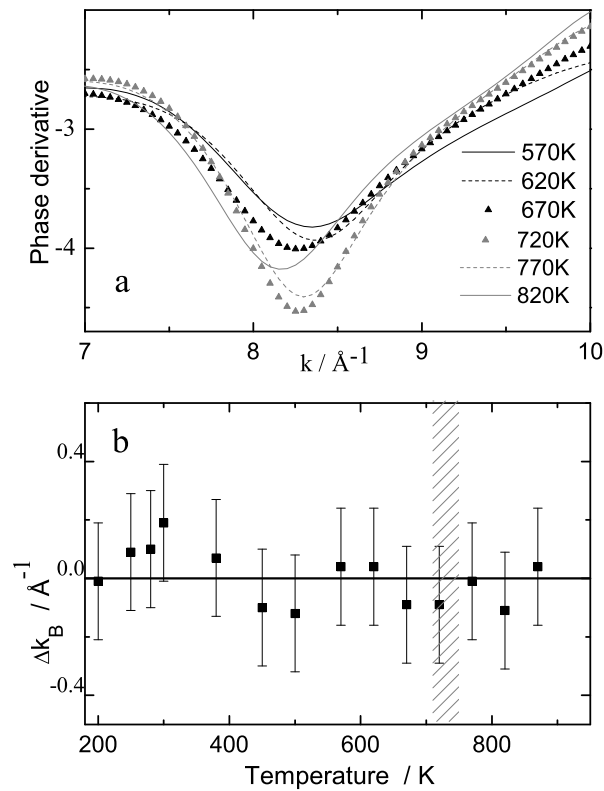}

\caption{\label{fig5-beat}a. Total phase shift derivative: the minimum of
the curve defines the beat position $k_{B}$. b. Variation of the
$k_{B}$ value with the temperature. The domain where orthorhombic
to cubic transition takes place is marked by the hatched area. }
\end{figure}

The $PD$ method was applied in the analysis of this coordination
shell contribution, first for the room temperature $(RT)$ spectrum,
where the local structure is already well established. The EXAFS signal
was back Fourier transformed in the R-range 0.7\AA~to 2.1\AA~
(fig.\ref{fig4-Phase}a). A minimum in the total amplitude (fig.\ref{fig4-Phase}a)
and an inflexion point in the total phase (fig.\ref{fig4-Phase}b)
are observed at $k_{B}=8.3$\AA$^{-1}$. This beat position corresponds
to a value $\delta R_{B}=\pi/2k_{B}=0.19$\AA, in good agreement
with bond length separation obtained from crystallographic data ($\delta$R=
0.21\AA). The small difference, most likely due to the approximations
involved in the $PD$ method\cite{Martens-PRL77}, is commonly reported
\cite{Jaffres-PRB00,Jiang-PRB91} and does not affect the precision
in the relative modification of this separation, that should be revealed
by a shift in the beat position. Identical procedure of the phase
extraction have been applied to the EXAFS spectra in the whole temperature
range, leading to a beat position value comprised between 8.1 and
8.5\AA$^{-1}$(Fig.\ref{fig5-beat}a). A relative error of 5\% has
been estimated from the mean square deviation for spectra collected
at the same temperature. Within this error bar the same value of k$_{B}$
is obtained over the whole temperature range, showing that the bond
length splitting is kept constant across $T^{*}$~(Fig.\ref{fig5-beat}b).

\begin{figure}
\includegraphics{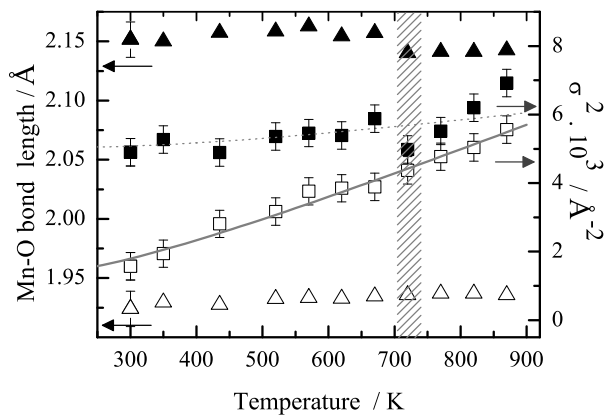}

\caption{\label{fig6-distsigma}Mn-O bond lengths (triangles, left scale)
and associated DWF (square, right scale). The open symbols and plain
symbols correspond to short and long lengths respectively. For seek
of clarity the error bars in the bond lengths, almost constant over
the temperature range, have been shown only for room temperature.
The plain line is the fit of the thermal disorder using the correlated
Einstein model. The dot line is only a guide for the eyes. A small
drop of the long bond DWF is observed at $T^{*}$(hatched area). At
this temperature the DWF of the short and long bonds are equalized. }
\end{figure}

A two-shell fit in k-space is then performed using theoretical amplitude
and phase functions generated using FEFF7 code \cite{Rehr-PRL92,Rehr-JAChemSoc91}
from the crystallographic \textit{Pbnm} structure of LaMnO$_{3}$.
The number of first neighbors were fixed ($N_{1}=4$ for the short
$(Mn-O)_{s}$ bonds and $N_{2}=2$ for the $(Mn-O)_{l}$ long bonds)
and the value of the amplitude reduction factor $S_{0}^{2}$ was determined
from $RT$ spectra and then fixed for the analysis of the spectra
at other temperatures. Figure \ref{fig6-distsigma} shows the results
of the fit. We note that the $Mn-O$~distances remain almost constant
through $T^{*}$ at values close to those known for LaMnO$_{3}$~at
$RT$ : $(Mn-O)_{s}=1.93$\AA~and $(Mn-O)_{l}=2.15$\AA, confirming
that there is no collapse of these distances across the transition. 

As concerns the disorder, the study of the Debye-Waller term brings
about some interesting observations. As already been pointed out in
previous XAS studies \cite{Shibata-PRB03,Subias-PRB98}, the disorder
associated to the long and the short bonds at room temperature are
quite different. According to the prediction of Goodenough \cite{Goodenough-PR55}
about the character of the Mn-O bonding, the short $(Mn-O)_{s}$~bonds
are stable bonds with semicovalent character, while the long $(Mn-O)_{l}$
distance corresponds to a weaker ionic bonding. The static radial
disorder associated to the ionic long bonds is large ($\sigma=0.07$
\AA) and keeps almost the same value over the whole temperature range
below $T^{*}$ . The radial disorder associated to the short bonds
is about three times smaller at room temperature, in agreement with
the results reported in the literature. The evolution of this factor
is thermally driven, well accounted by a simple correlated model for
the thermal vibrations\cite{Sevillano-PRB79}. The temperature found
for the Einstein model is $\Theta_{E}\sim680K$ (Fig.\ref{fig6-distsigma}),
in agreement with the high Debye temperatures ($\Theta_{D}\sim3/2\Theta_{E}\sim1000K$)
already reported in perovskites compounds\cite{Proffen-PRB99,Bridges-PRB00}.
The DWF associated to long bond is almost temperature independent
up to values close to $T^{*}$. At this temperature we observe a small
drop in this factor, partially screened by the error on the determination
of $\sigma^{2}$. We should point out that at this temperature the
DWF associated to short and long bonds assume essentially the same
values. Above $T^{*}$ both factors seem to follow a simple thermal
behavior, in the continuity of the low temperature behavior of the
short bond DWF. 

The small drop in the long bond DWF is consistent with the small discontinuity
in the long bond -related side in the FT (Fig.\ref{fig3-FT}). As
the distances are kept constant and the DWF associated to the short
bond increases continuously, it should be related to the drop in the
height of the peak in the FT across the transition. We should emphasize
that the DWF in EXAFS only accounts for radial disorder, \textit{i.e.}
the relative variation in the Mn and O positions,  projected along
the bond axis. In the basal plane, all the Mn-O-Mn linkages are composed
of a long and a short bond. The longitudinal variation of $(Mn-O)_{l}$
length should then be accommodated by flattening and unbending the
Mn-O-Mn angle. The resulting transverse accommodation has a limited
projection on the $(Mn-O)_{s}$~ bond, and is not {}``seen'' in
the radial disorder term. Additionally, the accommodation through
the Mn-O-Mn angle yield to destructive interferences in the different
multiple scattering contributions of the almost linear 3-leg MS paths
(Fig.\ref{fig2-EXAFS}a- insert) and limits the total contribution
to the EXAFS signal. Around $T^{*}$ ~the radial DWF of $(Mn-O)_{s}$~bonds
becomes of the same order as that of $(Mn-O)_{l}$ bonds and the variations
of the short and long bonds are almost decorrelated. The two terms
follow then similar thermal behaviors. The Mn-O-Mn angle is no more
so strongly involved in the accomodation the bond vibrations and the
contribution of the multiple scattering path involving the two adjacent
Mn and their common O increases. The reduction of the angular distortion
favors an increase of the hybridization between the Mn 3d and O 2p
orbitals. 

\begin{figure}
\includegraphics{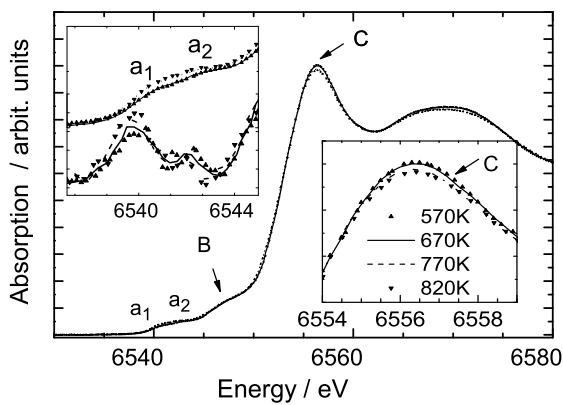}

\caption{\label{fig1-XANES} XANES spectra across the orbital ordering temperature
($\sim710K$). The insets are close views showing the subtle discontinuities
in amplitude. Left inset : preedge region and corresponding absorption
derivative. Right inset: main feature C. }
\end{figure}

The XANES spectra across the transition (570K to 820K) are shown in
Fig.\ref{fig1-XANES}. The spectra exhibit the features already reported
in LaMnO$_{3}$ \cite{Croft-PRB97,Bridges-PRB00,Bridges-PRB01a,Qian-PRB03,Qian-PRB00}.
The two pre-edge peaks $a_{1}$ and $a_{2}$ \textcolor{black}{(6540.0
and 6542.4eV) arise essentially from 1s-4p dipole allowed transitions
of Mn $3d$ character. They account} for the hybridization of the
Mn $4p$ orbitals with the Mn \textcolor{black}{$3d$} orbitals on
neighboring Mn ions, either directly or via O atoms \cite{Elfimov-PRL99,Hozoi-PRB01}.
In reduced symmetry due to the local Jahn-Teller distortion of the
MnO$_{6}$ octahedra, the $1s\rightarrow4p$ transition on Mn site
is split into separable components close to the onset (shoulder B)
around 6546eV and at the top of the rising edge (C main peak) in the
region 6550-6560eV where lies most of the spectral strength \cite{Bridges-PRB00}.
There is no energy shift in the XANES over the whole temperature range,
in agreement with the absence of significant modification in the average
coordination distance. It has been recently invoked \cite{Zhou-PRB03}
that a charge disproportionation for the Mn ions could happen at the
transition temperature. However this would result in a splitting in
the rising edge that has not been observed. Drastic changes in the
charge disproportionation should then be excluded. The top of the
main feature C measures the net $4p-$hole count. Up to the transition
temperature the intensity of this C peak decreases very slowly and
continuously when the temperature increases as a consequence of the
increase of the thermal disorder. At the transition a small additional
drop in the maximum peak height takes place (Fig.\ref{fig1-XANES},
inset). We should remind that the experimental procedure based on
the collection of the XAS data twice, both at increasing and decreasing
temperature, assure us the perfect reproducibility of the XANES features.
Small changes in the intensity of the features, especially in the
maximum peak height, are not experimental artifacts. An additional
drop in the hole count is then observed across the transition. It
accounts for an additional non-thermal increase of the delocalization
of the $e_{g}$ electrons, through a band effect associated to the
increase of the average Mn-O bond covalence\cite{Elfimov-PRL99}.
Correlatively, subtle but reproducible modifications are observed
in the preedge domain (Fig.\ref{fig1-XANES} inset). Alterations in
the spectral weight of these features are currently associated to
modification in the distribution of $e_{g}$ majority and ($e_{g}$
$t_{2g}$) minority states \cite{Bridges-PRB00}. In the present case
we do not observe spectral transfer from one to the other features,
but the simultaneous enhancement of the two features. We associate
this enhancement to an increase of the hybridization between the Mn
$4p$ orbitals and the Mn $3d$ orbitals on neighboring atoms. This
is consistent with a reduction of the disorder in the superexchange
angle deduced from EXAFS analysis. The delocalization of the $e_{g}$
electrons due to the thermal disorder partly overcome the localization
associated to the ionic bonds and should be related to the reported
drop in the resistivity at the orbital ordering temperature\cite{Zhou-PRB99,Qian-PRB03}.

\section{Conclusions}

We reported a XAS study of the local atomic and structural modifications
around the manganese atoms in LaMnO$_{3}$ across the orbital ordering
transition $T^{*}$. The Jahn-Teller splitting into long $(Mn-O)_{l}$
and short $(Mn-O)_{s}$ bonds within the MnO$_{6}$ octahedra is kept
in the high temperature phase. We observe that significant electronic
and structural changes are taking place across this transition at
the Mn site. We show that the structural modifications correspond
to a change in the accommodation of the local thermal disorder related
to vibrations of the $Mn-O$ bonds. We point out that the orbital
ordering transition takes place when the relative displacement of
the atoms along the long and short bonds becomes of the same order.
Below $T^{*}$, the radial thermal disorder appears to be mostly accomodated
by bending of the tilt angle Mn-O-Mn among adjacent octahedra. Above
$T^{*}$, the Mn-O-Mn angle are flattened and strengthened and the
average radial disorder becomes isotropic, allowing a complete decorrelation
of the bond length variations. Subtle electronic modifications at
the Mn sites also take place : a small drop in the 4p counts and a
small enhancement of the hybridization between 4p and 3d orbitals
are observed across the transition. These modifications are associated
to an increase of the average covalent character of the $Mn-O$ bonds.

This work is partially supported by LNLS/ABTLuS/MCT. RAS acknowledges
the PIBIC/CNPq. AYR acknowledges the grant from CNPq/PCI

\end{document}